\documentclass[12pt]{article}

\usepackage[authoryear]{natbib}
\usepackage{graphicx}
\usepackage{amssymb}
\usepackage{amsmath}
\usepackage{times}
\usepackage{bm}

\newcommand{\diag}{{\rm diag}}
\newcommand{\tr}{{\rm tr}}
\newcommand{\Es}{{\rm E}}

\newcommand{\BS}{{\rm BS}}
\newcommand{\EBS}{{\rm EBS}}
\newcommand{\SHN}{{\rm SHN}}
\newcommand{\SSN}{{\rm SSN}}
\newcommand{\SN}{{\rm SN}}

\title{A log-Birnbaum--Saunders Regression Model with Asymmetric Errors}
\author{Artur J.~Lemonte\\
{\small {\em Departamento de Estat\'istica, Universidade de S\~ao Paulo,
S\~ao Paulo/SP, 05508-090, Brazil}}}
\date{}

\begin{document}
\maketitle

\begin{abstract}

The paper by \cite{Leiva-et-al-2010-CSTM} introduced a skewed version of the sinh-normal
distribution, discussed some of its properties and characterized an extension
of the Birnbaum--Saunders distribution associated with this distribution.
In this paper, we introduce a skewed log-Birnbaum--Saunders regression model
based on the skewed sinh-normal distribution.
Some influence methods, such as the local influence and generalized leverage
are presented. Additionally, we derived the normal curvatures of local
influence under some perturbation schemes. An empirical application to a
real data set is presented in order to illustrate the usefulness of the
proposed model.\\


\end{abstract}

\section{Introduction}\label{introduction}

The two-parameter Birnbaum-Saunders (BS) distribution, also known as the fatigue life distribution,
was introduced by \cite{BSa1969a, BSa1969b}.
It was originally derived from a model for a physical fatigue process where dominant
crack growth causes failure. A more general derivation was provided by \cite{Desmond1985}
based on a biological model and relaxing
several of the assumptions made by \cite{BSa1969a}. \cite{Desmond1986} investigated
the relationship between the BS distribution and the
inverse Gaussian distribution. The author established that the
BS distribution can be written as a mixture equally weighted
from an inverse Gaussian distribution and its complementary reciprocal.

The random variable $T$ is said to have a BS distribution with
parameters $\alpha, \eta > 0$, say $\BS(\alpha, \eta)$,
if its cumulative distribution function (cdf) is given by
$F(t)=\Phi(v)$, $t > 0$, where $\Phi(\cdot)$ is the standard normal distribution function,
$v=\rho(t/\eta)/\alpha$, $\rho(z)= z^{1/2}-z^{-1/2}$
and $\alpha$ and $\eta$ are shape and scale parameters,
respectively. Also, $\eta$  is the median of the distribution: $F(\eta) = \Phi(0) = 1/2$.
For any constant $k > 0$, it follows that $kT \sim\BS(\alpha, k\eta)$.
It is noteworthy that the reciprocal property holds for the BS
distribution: $T^{-1} \sim\BS(\alpha, \eta^{-1})$; see
\cite{Saunders1974}. The BS distribution has received considerable attention
over the last few years. \cite{kundu-et-al-2008} discussed the shape of the hazard
function of the BS distribution. Results on improved statistical
inference for the BS distribution are discussed in \cite{WuWong2004} and \cite{LCNV07, LSCN08}.
Some generalizations and extensions of
the BS distribution are presented in \cite{Diaz-Leiva05},
\cite{GPB09}, \cite{guiraud-et-al-2009} and \cite{Castillo-et-al-2009}.
This distribution has been applied in reliability studies
\citep[see, for example,][]{Balakrishnan-et-al-2007} and
outside this field; see \cite{Leiva-et-al-2008} and \cite{Leiva-et-al-2009}.
Additionally, based on the BS distribution, \cite{Bhatti2010}
introduced the BS autoregressive conditional duration model.
\cite{XiTang10} presented estimators for the unknown parameters of the
BS distribution using reference prior.

From \cite{Rieck89}, if
\begin{equation}\label{rv1}
Z = \nu + \frac{2}{\alpha}\sinh\biggl(\frac{Y - \gamma}{\sigma}\biggr)\sim{\rm N}(0, 1),
\end{equation}
then $Y$ has a four-parameter sinh-normal (SHN) distribution, denoted
by $Y\sim\SHN(\alpha, \gamma, \sigma,\nu)$, where $\nu\in\Re$ and $\alpha>0$
are the shape parameters, and $\gamma\in\Re$ and $\sigma>0$
correspond to the location and scale parameters, respectively.
According to \cite{Rieck89}, the parameter
$\nu$ is also the noncentralty parameter. If $\nu=0$,
the notation is reduced simply by $Y\sim\SHN(\alpha, \gamma, \sigma)$, and
this distribution has a number of interesting properties.
For example, it is symmetric around the mean $\Es(Y)=\gamma$,
it is unimodal for $\alpha\leq 2$ and bimodal for $\alpha > 2$ and if
$Y_{\alpha}\sim\SHN(\alpha,\gamma,\sigma)$, then
$Z_{\alpha}=2(Y_{\alpha}-\gamma)/(\alpha\sigma)$
converges in distribution to the standard normal distribution when $\alpha\to 0$.
If $Y\sim\SHN(\alpha, \gamma, \sigma=2)$, then $T=\exp(Y)$ follows the
BS distribution with shape parameter $\alpha$ and scale parameter $\eta=\exp(\gamma)$,
i.e.~$T=\exp(Y)\sim\BS(\alpha,\eta)$. For this reason,
according to \cite{Leiva-et-al-2010-CSTM}, the SHN distribution
is also called the log-Birnbaum--Saunders (log-BS) distribution.
Additionally, according to these authors, the SHN and BS models
corresponding to a logarithmic distribution and its associated
distribution, respectively \citep[][Ch.~12]{MO2007}.

\cite{RiekNedelman91} introduced a log-BS regression model based on the
$\SHN(\alpha,\gamma,2)$ distribution. Their
regression model has been studied by several
authors. Some important references are \cite{Tisionas01}, \cite{Galea-etal-2004},
\cite{LBPG2007}, \cite{Desmond-et-al-2008}, \cite{Lemonte-et-al-2010},
\cite{Xiao-et-al-2010} and \cite{Cancho-et-al-2010}, among others.
Generalizations of the log-BS regression model introduced by \cite{RiekNedelman91} are
presented in \citet[][\S~4]{XiWei07} and \cite{LemCord09}.

\cite{Leiva-et-al-2010-CSTM} introduced a skewed SHN
distribution by replacing the standard normal distribution in
equation (\ref{rv1}) by the skew-normal (SN) distribution \citep{Azzaline1985}, i.e.~they
consider the random variable
\[
Z = \nu + \frac{2}{\alpha}\sinh\biggl(\frac{Y - \gamma}{\sigma}\biggr)\sim\SN(\lambda),
\]
where $\lambda\in\Re$ is the shape parameter which determines the skewness.
Now, the notation used is $Y\sim\SSN(\alpha,\gamma,\sigma,\nu,\lambda)$.
From now on, we shall consider $\nu=0$ and $\sigma = 2$ and hence the notation is
given by $Y\sim\SSN(\alpha,\gamma,2,\lambda)$. The random variable
$T=\exp(Y)$ follows the extended Birnbaum--Saundres (EBS) distribution,
with shape parameters $\alpha>0$ and $\lambda\in\Re$, and scale parameter
$\eta=\exp(\gamma)$. Now, the notation is $T=\exp(Y)\sim\EBS(\alpha, \eta,\lambda)$.

\begin{figure}
\centering
\includegraphics[scale=0.385]{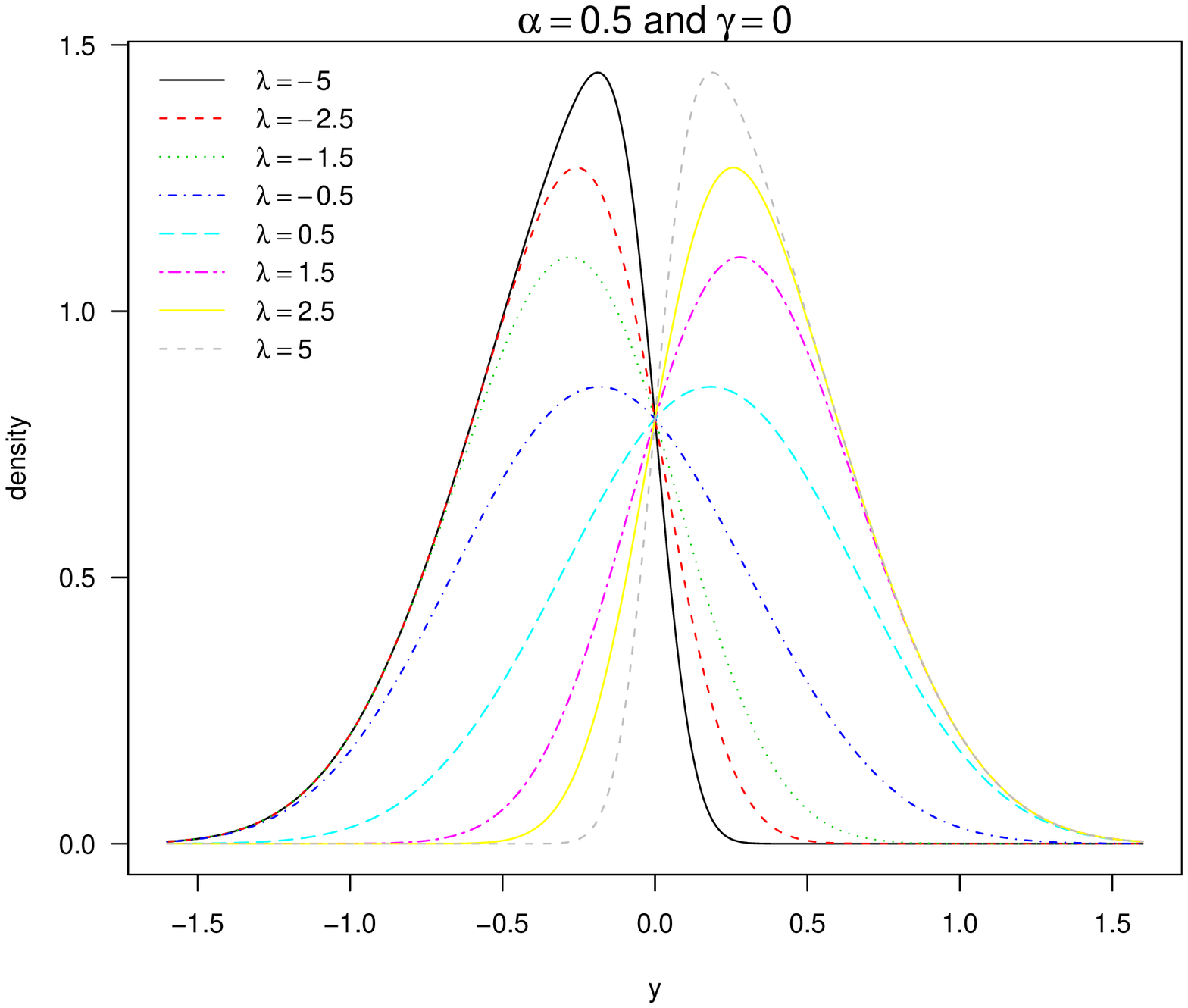}
\includegraphics[scale=0.385]{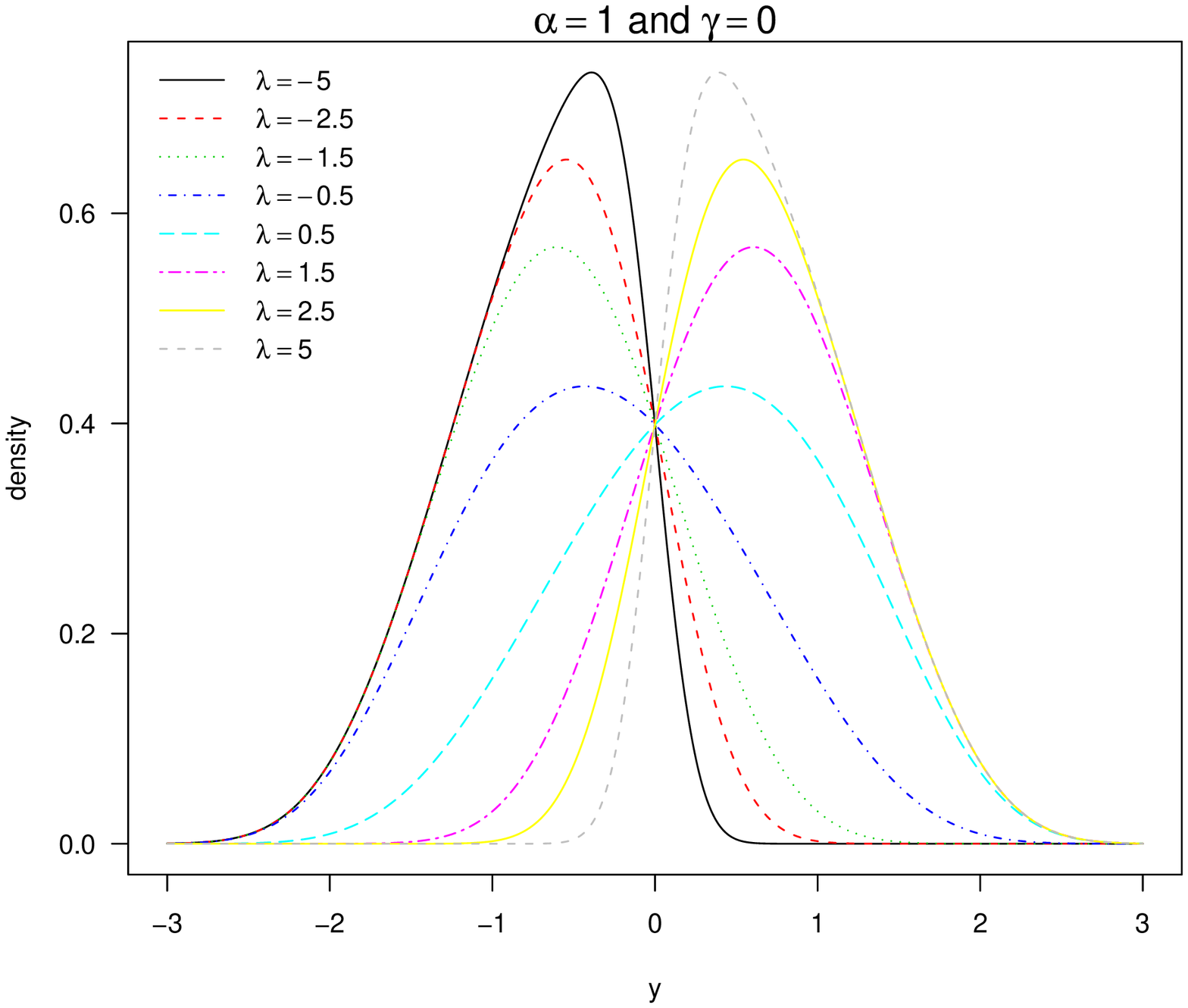}
\includegraphics[scale=0.385]{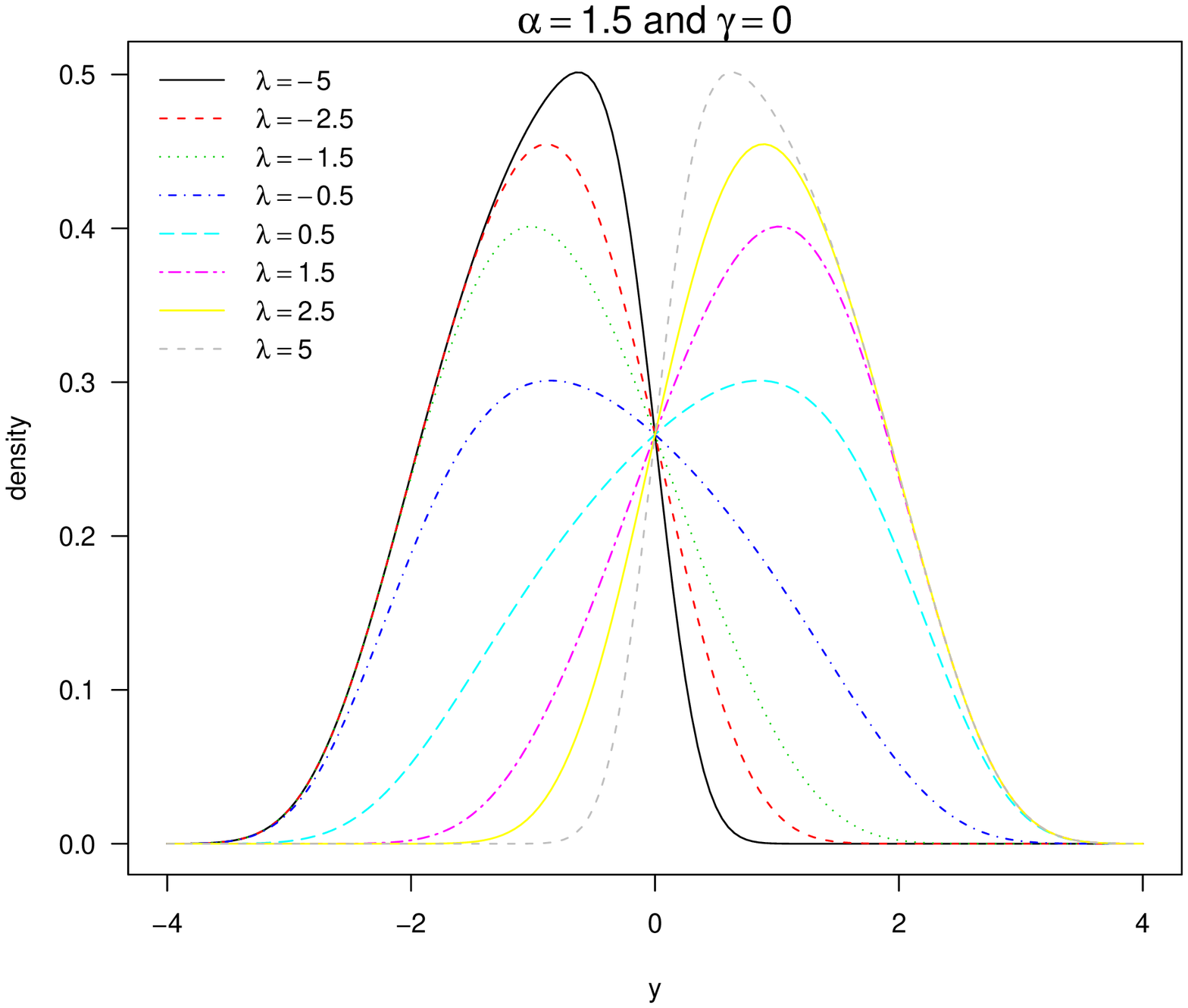}
\includegraphics[scale=0.385]{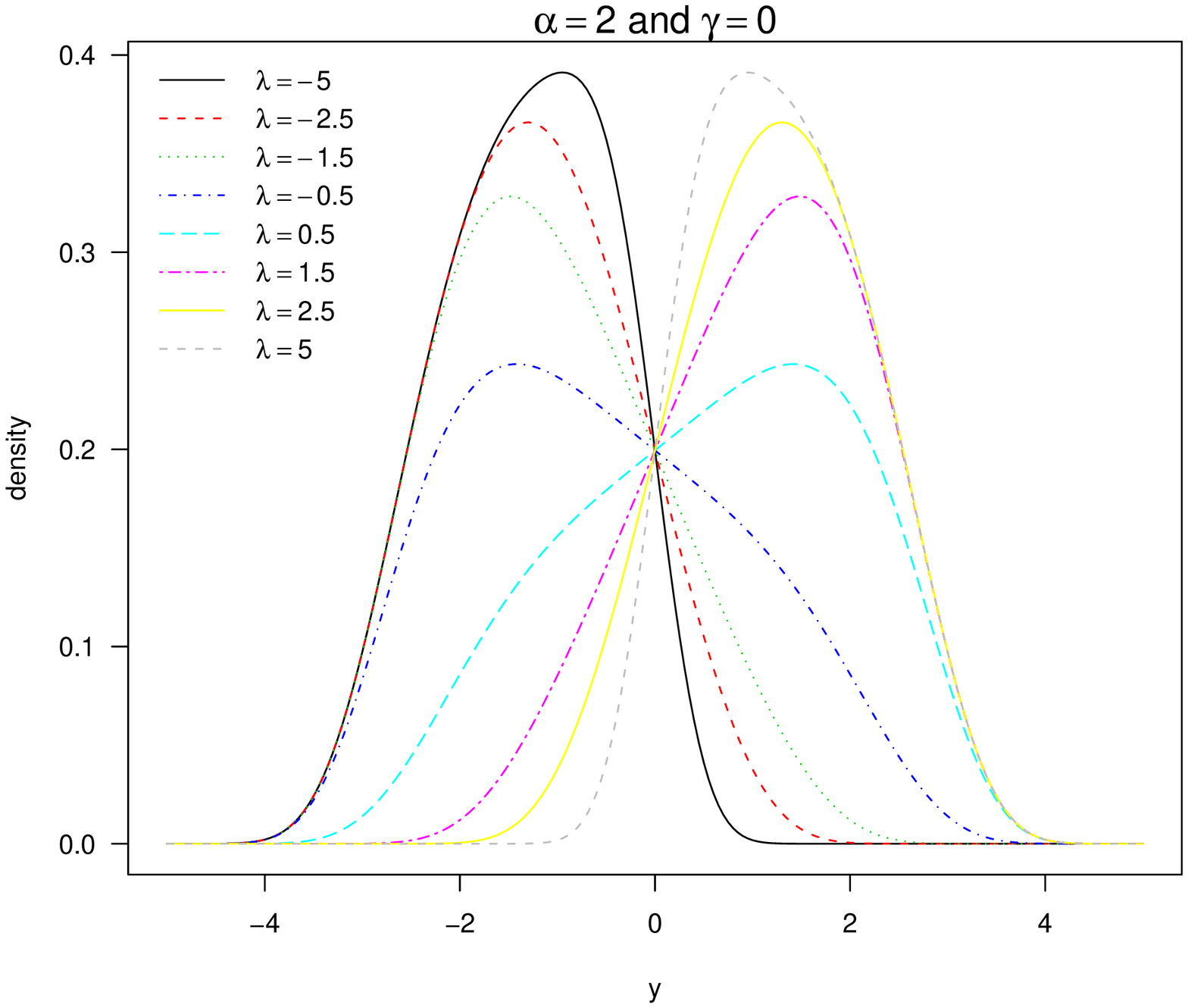}
\includegraphics[scale=0.385]{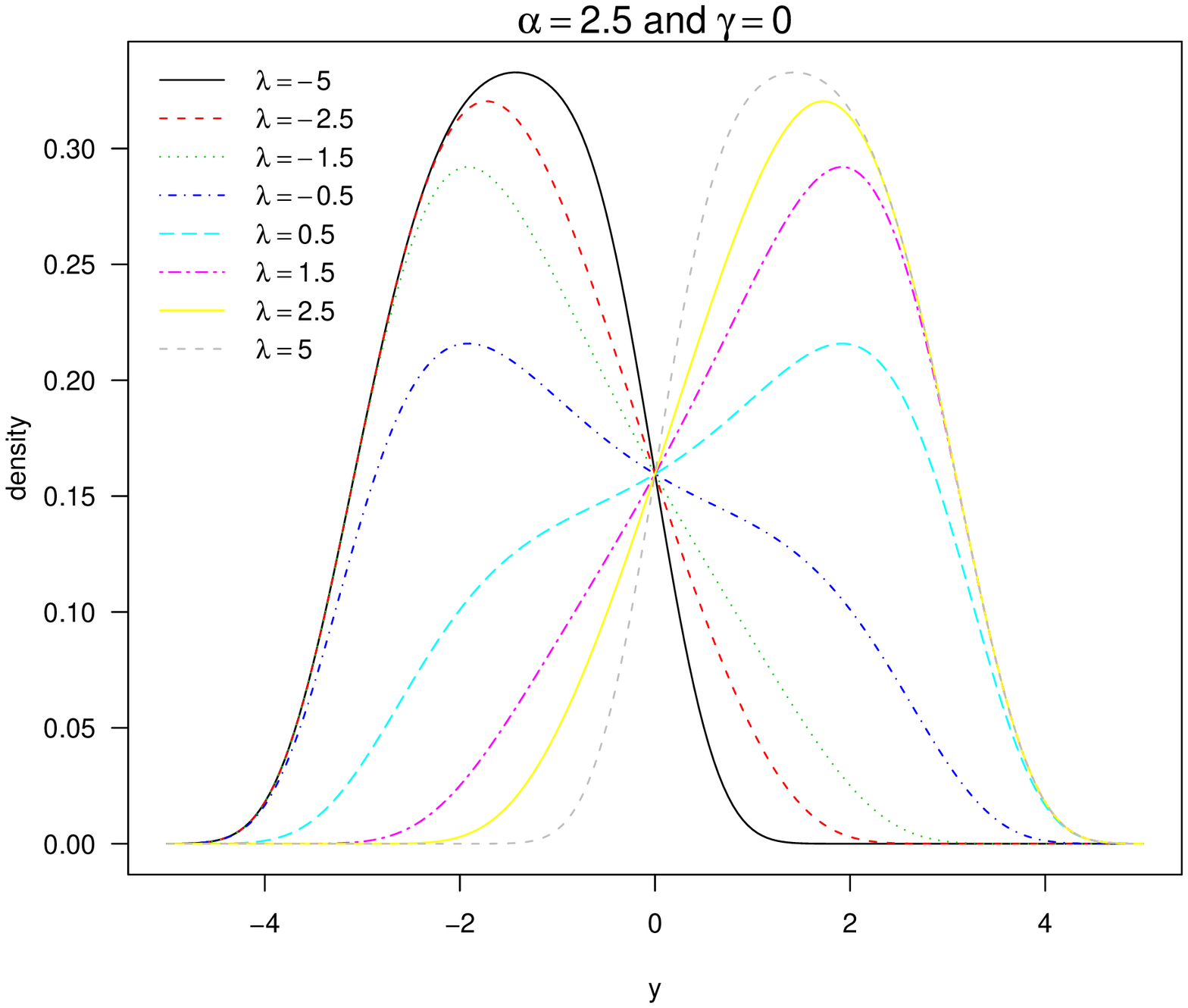}
\includegraphics[scale=0.385]{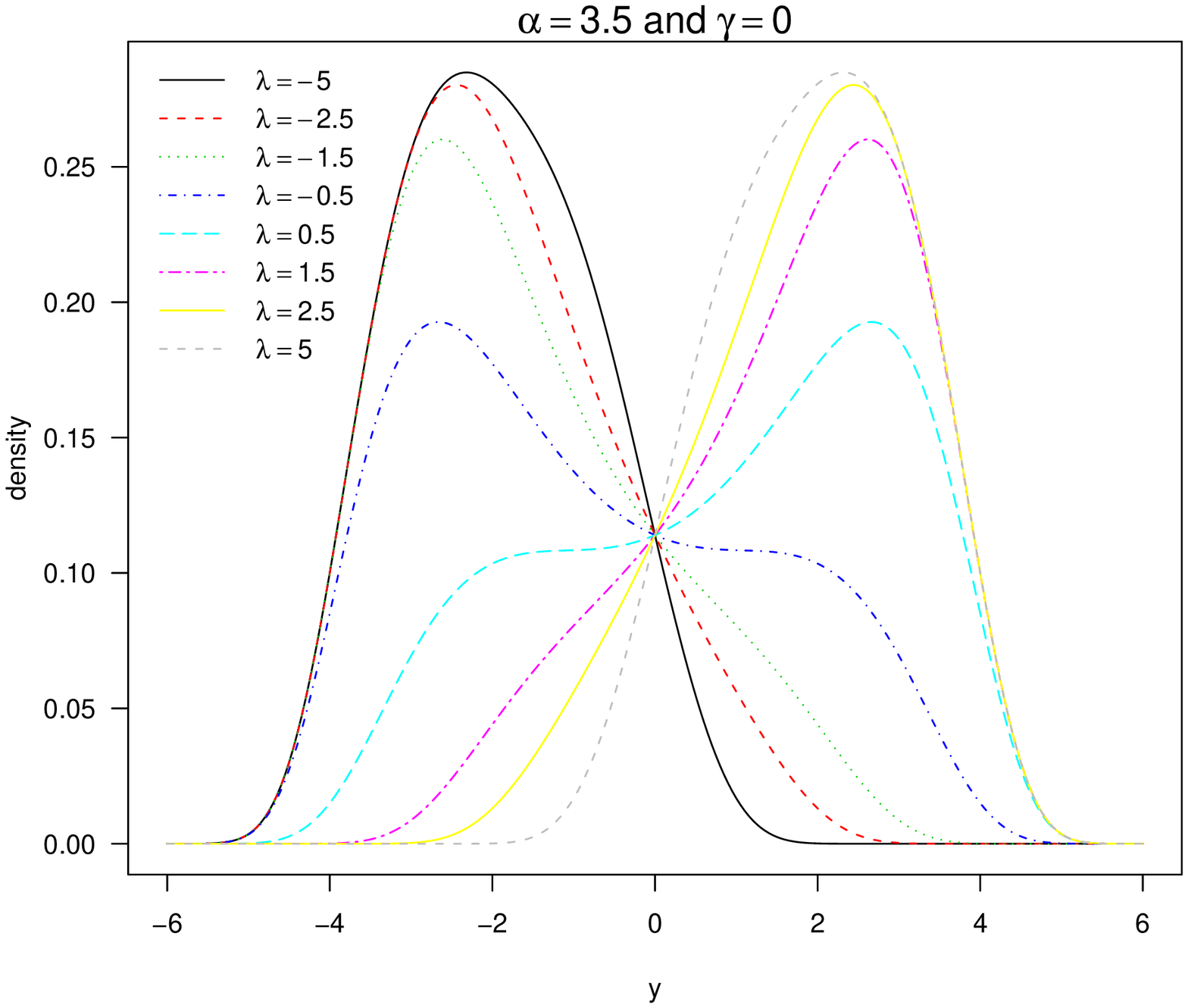}
\caption{Plots of the density function of the SSN
distribution for some parameter values.}\label{densities_plots}
\end{figure}

Let $T\sim\EBS(\alpha, \eta,\lambda)$. The
density function of $Y=\log(T)$ is given by \citep{Leiva-et-al-2010-CSTM}
\[
\pi(y)=\frac{2}{\alpha}\cosh\biggl(\frac{y-\gamma}{2}\biggr)
\phi\biggl(\frac{2}{\alpha}\sinh\biggl(\frac{y-\gamma}{2}\biggr)\biggr)
\Phi\biggl(\frac{2\lambda}{\alpha}\sinh\biggl(\frac{y-\gamma}{2}\biggr)\biggr),
\qquad y\in\Re,
\]
where $\phi(\cdot)$ is the standard normal density function,
and, as before, we write $Y\sim\SSN(\alpha,\gamma,2,\lambda)$.
The $s$th ($s=1,2,\ldots$) moment of $Y$ can be written as
\[
\Es(Y^s) = 2^k\sum_{k=0}^{s}\gamma^{s-k}c_{k}(\alpha,\lambda),\qquad
c_{k}(\alpha,\lambda) = \int_{-\infty}^{\infty}\{\sinh^{-1}(\alpha w/2)\}^{k}\phi(w)\Phi(\lambda w)dw.
\]
Thus, the mean of $Y$ is given by $\Es(Y) = \gamma + c(\alpha,\lambda)$, with
\[
c(\alpha,\lambda) = 4\int_{-\infty}^{\infty}\{\sinh^{-1}(\alpha w/2)\}\phi(w)\Phi(\lambda w)dw.
\]
Plots of the $\SSN(\alpha,\gamma,2,\lambda)$ distribution are
illustrated in Figure \ref{densities_plots} for selected
parameter values.

The chief goal of this paper is to introduce a skewed log-BS regression model based on
the $\SSN(\alpha,\gamma,2,\lambda)$ distribution, recently proposed by \cite{Leiva-et-al-2010-CSTM}.
The proposed regression model is convenient for modeling asymmetric
data, and it is an alternative to the log-BS regression model introduced
by \cite{RiekNedelman91} when the data present skewness.
The article is organized as follows. Section~\ref{model}
introduces the class of skewed log-BS regression
models. The score functions and observed information matrix are given.
Section \ref{local_influence} deals with some
basic calculations related with local influence.
Derivations of the normal curvature under different perturbation schemes
are presented in Section~\ref{curvatures}. Generalized leverage
is derived in Section \ref{leverage}.
Section~\ref{application} contains an application to a real data set of
the proposed regression model.
Finally, concluding remarks are offered in Section~\ref{conclusion}.

\section{Model specification}\label{model}

The skewed log-BS regression model is defined by
\begin{equation}\label{eq1}
y_{i} = \bm{x}_{i}^\top\bm{\beta} + \varepsilon_{i},\qquad i=1\ldots,n,
\end{equation}
where $y_i$ is the logarithm of the $i$th observed lifetime,
$\bm{x}_{i} = (x_{i1},\ldots,x_{ip})^\top$ is a vector of known explanatory variables
associated with the $i$th observable response $y_{i}$,
$\bm{\beta}=(\beta_1,\ldots,\beta_p)^{\top}$ is a vector of
unknown parameters, and the random errors
$\varepsilon_{i}\sim{\rm SSN}(\alpha,-c(\alpha,\lambda),2,\lambda)$ that corresponds
to the regression model where the error distribution has mean zero. Thus,
we have $y_{i}\sim{\rm SSN}(\alpha,\bm{x}_{i}^\top\bm{\beta}-c(\alpha,\lambda),2,\lambda)$,
with $\Es(y_{i}) = \bm{x}_{i}^\top\bm{\beta}$, for $i=1,\ldots,n$.

The log-likelihood function for the vector parameter
$\bm{\theta}=(\bm{\beta}^{\top},\alpha, \lambda)^{\top}$
from a random sample $\bm{y}=(y_1,\ldots,y_n)^{\top}$ obtained from~(\ref{eq1})
can be expressed as
\begin{equation}\label{eq2}
\ell(\bm{\theta})=\sum_{i=1}^{n}\ell_{i}(\bm{\theta}),
\end{equation}
where $\ell_{i}(\bm{\theta}) = -\log(2\pi)/2 + \log(\xi_{i1}) - \xi_{i2}^{2}/2 +
\log\{\Phi(\lambda\xi_{i2})\}$,
\begin{equation*}\label{xis}
\xi_{i1}=\xi_{i1}(\bm{\theta}) = \frac{2}{\alpha}
\cosh\biggl(\frac{y_i - \bm{x}_{i}^\top\bm{\beta} + c(\alpha,\lambda)}{2}\biggr),
\quad
\xi_{i2}=\xi_{i2}(\bm{\theta}) = \frac{2}{\alpha}
\sinh\biggl(\frac{y_i - \bm{x}_{i}^\top\bm{\beta} + c(\alpha,\lambda)}{2}\biggr),
\end{equation*}
for $i=1,\ldots,n$. The function $\ell(\bm{\theta})$ is assumed to be
regular \citep[][Ch.~9]{CoxHinkley1974} with respect to all $\bm{\beta}$,
$\alpha$ and $\lambda$ derivatives up to second order.
Further, the $n\times p$ matrix $\bm{X}=(\bm{x}_{1}, \ldots, \bm{x}_{n})^{\top}$
is assumed to be of full rank, i.e., rank($\bm{X})=p$.

By taking the partial derivatives of the log-likelihood function with respect to
$\bm{\beta}$, $\alpha$ and $\lambda$, we obtain the components of the score vector
$\bm{U}_{\bm{\theta}} = (\bm{U}_{\bm{\beta}}^\top, U_{\alpha}, U_{\lambda})^\top$. We have
$\bm{U}_{\bm{\beta}} = \bm{X}^{\top}\bm{s}$, where $\bm{s}=(s_1,\ldots,s_n)^\top$
with $s_{i} = \{\xi_{i1}\xi_{i2}-\xi_{i2}/\xi_{i1} - \lambda\xi_{i1}\phi(\lambda\xi_{i1})/\Phi(\lambda\xi_{i2})\}/2$,
\begin{align*}
U_{\alpha} &= -\frac{n}{\alpha} + \frac{1}{\alpha}\sum_{i=1}^{n}\xi_{i2}^{2} - \frac{c_{\alpha}}{2}
\sum_{i=1}^{n}\biggl\{\xi_{i1}\xi_{i2}-\frac{\xi_{i2}}{\xi_{i1}}\biggr\}\\
&\quad+ \lambda\sum_{i=1}^{n}\frac{\phi(\lambda\xi_{i2})}{\Phi(\lambda\xi_{i2})}
\biggl\{\frac{c_{\alpha}\xi_{i1}}{2} - \frac{\xi_{i2}}{\alpha}\biggr\},
\end{align*}
\[
U_{\lambda} = -\frac{c_{\lambda}}{2}\sum_{i=1}^{n}\biggl\{\xi_{i1}\xi_{i2}-\frac{\xi_{i2}}{\xi_{i1}}\biggr\}
+  \frac{1}{2}\sum_{i=1}^{n}\frac{\phi(\lambda\xi_{i2})}{\Phi(\lambda\xi_{i2})}
\{\lambda c_{\lambda}\xi_{i1} +2\xi_{i2}\},
\]
where
\[
c_{\alpha} = c_{\alpha}(\alpha,\lambda) = 4\int_{-\infty}^{\infty}w(4+\alpha^2w^2)^{-1/2}\phi(w)\Phi(\lambda w)dw,
\]
\[
c_{\lambda} =c_{\lambda}(\alpha,\lambda) = 4\int_{-\infty}^{\infty}w\sinh^{-1}(\alpha w/2)\phi(w)\Phi(\lambda w)dw.
\]
Setting these equations to zero, $\bm{U}_{\bm{\theta}}=\bm{0}$, and
solving them simultaneously yields the MLE
$\widehat{\bm{\theta}} = (\widehat{\bm{\beta}}^\top,\widehat{\alpha},\widehat{\lambda})^\top$
of $\bm{\theta} = (\bm{\beta}^\top,\alpha,\lambda)^\top$. These equations cannot be
solved analytically and statistical software can be used to solve them numerically.
For example, the BFGS method \citep[see,][]{NocedalWright1999,
Press-et-al-2007} with analytical derivatives can be used
for maximizing the log-likelihood function $\ell(\bm{\theta})$.
Starting values $\bm{\beta}^{(0)}$, $\alpha^{(0)}$ and $\lambda^{(0)}$ are required.
Our suggestion is to use as an initial point
estimate for $\bm{\beta}$ the ordinary least squares estimate of this parameter vector,
that is, $\bar{\bm{\beta}} = (\bm{X}^{\top}\bm{X})^{-1}\bm{X}^{\top}\bm{y}$.
The initial guess for $\alpha$ we suggest is $\sqrt{\bar{\alpha}^2}$,
where
\[
\bar{\alpha}^2 = \frac{4}{n}\sum_{i=1}^{n}\sinh^{2}\biggr(\frac{y_{i} -
                        \bm{x}_{i}^{\top}\bar{\bm{\beta}}}{2}\biggl).
\]
We suggest $\lambda^{(0)} = 0$.
These initial guesses worked well in the application described in Section \ref{application}.

The asymptotic inference for the parameter vector
$\bm{\theta} = (\bm{\beta}^{\top},\alpha,\lambda)^{\top}$
can be based on the normal approximation of the MLE of $\bm{\theta}$,
$\widehat{\bm{\theta}} = (\widehat{\bm{\beta}}^\top,\widehat{\alpha},\widehat{\lambda})^\top$.
Under some regular
conditions stated in \citet[Ch.~9]{CoxHinkley1974} that are fulfilled for
the parameters in the interior of the parameter space,  we have
$\widehat{\bm{\theta}}\stackrel{a}{\sim}\mathcal{N}_{p+2}(\bm{\theta},\bm{\Sigma}_{\bm{\theta}})$,
for $n$ large, where $\stackrel{a}{\sim}$ means approximately
distributed and $\bm{\Sigma}_{\bm{\theta}}$ is the asymptotic variance-covariance matrix
for $\widehat{\bm{\theta}}$. The asymptotic behavior remains valid if $\bm{\Sigma}_{\bm{\theta}}$ is
approximated by $-\Ddot{\bm{L}}_{\widehat{\bm{\theta}}\widehat{\bm{\theta}}}^{-1}$, where
$-\Ddot{\bm{L}}_{\widehat{\bm{\theta}}\widehat{\bm{\theta}}}$ is the $(p + 2)\times(p + 2)$
observed information matrix evaluated at $\widehat{\bm{\theta}}$, obtained from
\[
\Ddot{\bm{L}}_{\bm{\theta}\bm{\theta}} =
\begin{bmatrix}
\Ddot{\bm{L}}_{\bm{\beta}\bm{\beta}} & \Ddot{\bm{L}}_{\bm{\beta}\alpha} &  \Ddot{\bm{L}}_{\bm{\beta}\lambda} \\
\Ddot{\bm{L}}_{\alpha\bm{\beta}} & \Ddot{L}_{\alpha\alpha}  & \Ddot{L}_{\alpha\lambda} \\
\Ddot{\bm{L}}_{\lambda\bm{\beta}} & \Ddot{L}_{\lambda\alpha} & \Ddot{L}_{\lambda\lambda}
\end{bmatrix}
=
\begin{bmatrix}
-\bm{X}^\top\bm{V}\bm{X} & -\bm{X}^\top\bm{h} &  -\bm{X}^\top\bm{b} \\
-\bm{h}^\top\bm{X} &  \tr(\bm{K}_1)  &  \tr(\bm{K}_2) \\
-\bm{b}^\top\bm{X} &  \tr(\bm{K}_2) &  \tr(\bm{K}_3)
\end{bmatrix},
\]
where
\[
\bm{V} = \diag\{v_1, \ldots,v_n\},\qquad
\bm{K}_{1} = \diag\{k_{i1},\ldots,k_{n1}\},\qquad
\bm{K}_{2} = \diag\{k_{i2},\ldots,k_{n2}\},
\]
\[
\bm{K}_{3} = \diag\{k_{i3},\ldots,k_{n3}\},\qquad
\bm{h} = (h_1,\ldots,h_n)^{\top},\qquad
\bm{b} = (b_1,\ldots,b_n)^{\top}.
\]
All the quantities necessary to obtain
the observed information matrix are given in the Appendix.

\section{Local influence}\label{local_influence}

The local influence method is recommended when the concern is
related to investigate the model sensibility
under some minor perturbations in the model (or data). Let
$\bm{\omega}\in \bm{\Omega}$ be a $k$-dimensional vector
of perturbations, where $\bm{\Omega}\subset\Re^k$
is an open set. The perturbed log-likelihood function
is denoted by $\ell(\bm{\theta}|\bm{\omega})$.
The vector of no perturbation is $\bm{\omega}_{0} \in \bm{\Omega}$, such that
$\ell(\bm{\theta}|\bm{\omega}_{0})=\ell(\bm{\theta})$.
 The influence of minor perturbations on the maximum likelihood estimate
$\widehat{\bm{\theta}}$ can be assessed by using the log-likelihood displacement
$LD_{\bm{\omega}} = 2\{\ell(\widehat{\bm{\theta}}) - \ell(\widehat{\bm{\theta}}_{\bm{\omega}})\}$,
where $\widehat{\bm{\theta}}_{\bm{\omega}}$ denotes the maximum likelihood
estimate under $\ell(\bm{\theta}|\bm{\omega})$.

The Cook's idea for assessing local influence is essentially
to analyse the local behavior of $LD_{\bm{\omega}} $
around $\bm{\omega}_{0}$ by evaluating the curvature of the plot of
$LD_{\bm{\omega}_{0} + a\bm{d}}$ against $a$, where $a\in\Re$ and $\bm{d}$ is a unit norm direction.
 One of the measures of particular interest is the direction $\bm{d}_{\max}$ corresponding to
the largest curvature $C_{\bm{d}_{\max}}$. The index plot of $\bm{d}_{\max}$ may evidence
those observations that have considerable influence on $LD_{\bm{\omega}} $ under minor
perturbations. Also, plots of $\bm{d}_{\max}$ against covariate
values may be helpful for identifying atypical
patterns. \cite{Cook1986} shows that the normal curvature
at the direction $\bm{d}$ is given by
\[
C_{\bm{d}}(\bm{\theta}) = 2|\bm{d}^{\top}\bm{\Delta}^{\top}
\Ddot{\bm{L}}_{\bm{\theta}\bm{\theta}}^{-1}\bm{\Delta}\bm{d}|,
\]
where $\bm{\Delta} = \partial^2\ell(\bm{\theta}|\bm{\omega})/
\partial\bm{\theta}\partial\bm{\omega}^{\top}$ and
$-\Ddot{\bm{L}}_{\bm{\theta}\bm{\theta}}$ is the observed information matrix,
both $\bm{\Delta}$ and $\Ddot{\bm{L}}_{\bm{\theta}\bm{\theta}}$ are
evaluated at $\widehat{\bm{\theta}}$ and $\bm{\omega}_{0}$.
Hence, $C_{\bm{d}_{\max}}/2$ is the largest eigenvalue of
$\bm{B} = -\bm{\Delta}^{\top}\Ddot{\bm{L}}_{\bm{\theta}
\bm{\theta}}^{-1}\bm{\Delta}$ and $\bm{d}_{\max}$
is the corresponding unit norm eigenvector. The index
plot of $\bm{d}_{\max}$ for the matrix
$\bm{B}$ may show how to perturb the model (or data)
to obtain large changes in the estimate of $\bm{\theta}$.

Assume that the parameter vector $\bm{\theta}$ is partitioned as
$\bm{\theta} = (\bm{\theta}_{1}^{\top},
\bm{\theta}_{2}^{\top})^{\top}$.  The dimensions of $\bm{\theta}_{1}$
and $\bm{\theta}_{2}$ are $p_{1}$ and $p - p_{1}$, respectively.  Let
\[
\ddot{\bm{L}}_{\bm{\theta}\bm{\theta}} =
\begin{bmatrix}
  \ddot{\bm{L}}_{\bm{\theta}_{1}\bm{\theta}_{1}} &
  \ddot{\bm{L}}_{\bm{\theta}_{1}\bm{\theta}_{2}} \\
  \ddot{\bm{L}}_{\bm{\theta}_{1}\bm{\theta}_{2}}^{\top} &
  \ddot{\bm{L}}_{\bm{\theta}_{2}\bm{\theta}_{2}}
\end{bmatrix},
\]
where $\ddot{\bm{L}}_{\bm{\theta}_{1}\bm{\theta}_{1}}
= \partial^2\ell(\bm{\theta})/
\partial\bm{\theta}_{1}\partial\bm{\theta}_{1}^{\top}$,
$\ddot{\bm{L}}_{\bm{\theta}_{1}\bm{\theta}_{2}}
= \partial^2\ell(\bm{\theta})/
\partial\bm{\theta}_{1}\partial\bm{\theta}_{2}^{\top}$ and
$\ddot{\bm{L}}_{\bm{\theta}_{2}\bm{\theta}_{2}}
= \partial^2\ell(\bm{\theta})/
\partial\bm{\theta}_{2}\partial\bm{\theta}_{2}^{\top}$. If
the interest lies on $\bm{\theta}_{1}$, the normal curvature in the
direction of the vector
$\bm{d}$ is $C_{\bm{d};\bm{\theta}_{1}}(\bm{\theta}) =
2|\bm{d}^{\top}\bm{\Delta}^{\top}(\ddot{\bm{L}}_{\bm{\theta}\bm{\theta}}^{-1}-
\ddot{\bm{L}}_{22})\bm{\Delta}\bm{d}|$, where
\[
\ddot{\bm{L}}_{22} =
\begin{bmatrix}
  \bm{0} & \bm{0} \\
  \bm{0} & \ddot{\bm{L}}_{\bm{\theta}_{2}\bm{\theta}_{2}}^{-1}
\end{bmatrix}
\]
and $\bm{d}_{\max;\bm{\theta}_{1}}$ here is the eigenvector
corresponding to the largest eigenvalue of $\bm{B}_{1} =
-\bm{\Delta}^{\top}(\ddot{\bm{L}}_{\bm{\theta}\bm{\theta}}^{-1} -
\ddot{\bm{L}}_{22})\bm{\Delta}$ \citep{Cook1986}.  The index plot of
the $\bm{d}_{\max;\bm{\theta}_{1}}$ may reveal those influential
elements on $\widehat{\bm{\theta}}_{1}$.

\section{Curvature calculations}\label{curvatures}

Next, we derive for three perturbation schemes the matrix
\[
\bm{\Delta} =
\frac{\partial^2\ell(\bm{\theta}|\bm{\omega})}
{\partial\bm{\theta}\partial\bm{\omega}^\top}
\biggr|_{\bm{\theta}=\widehat{\bm{\theta}},\bm{\omega}=\bm{\omega}_{0}}
=
\begin{bmatrix}
\bm{\Delta}_{\bm{\beta}} \\
\bm{\Delta}_{\alpha} \\
\bm{\Delta}_{\lambda}
\end{bmatrix},
\]
considering the model defined in~(\ref{eq1}) and its
log-likelihood function given by~(\ref{eq2}).
The quantities distinguished by the addition of ``\ $\widehat{}$\ ''
are evaluated at $\widehat{\bm{\theta}}=
(\widehat{\bm{\beta}}^\top, \widehat{\alpha},\widehat{\lambda})^\top$.

\subsection{Case-weights perturbation}

The perturbation of cases is done by defining some weights for
each observation in the log-likelihood function as follows:
\[
\ell(\bm{\theta}| \bm{\omega}) = \sum_{i=1}^n \omega_i\ell_{i}(\bm{\theta}),
\]
where $\bm{\omega} = (\omega_{1},\ldots,\omega_{n})^{\top}$ is the total vector of weights
and $\bm{\omega}_{0} = (1,\ldots,1)^{\top}$
is the vector of no perturbations.
After some algebra, we have
\[
\bm{\Delta}_{\bm{\beta}} =\bm{X}^{\top}\widehat{\bm{S}},
\qquad
\bm{\Delta}_{\alpha} = (\widehat{a}_{1},\ldots,\widehat{a}_{n}),
\qquad
\bm{\Delta}_{\lambda} = (\widehat{c}_{1},\ldots,\widehat{c}_{n}),
\]
where $\bm{S} = \diag\{s_1,\ldots, s_n\}$,
\[
a_{i} = -\frac{1}{\alpha} + \frac{\xi_{i2}^{2}}{\alpha} - \frac{c_{\alpha}}{2}
\biggl\{\xi_{i1}\xi_{i2} - \frac{\xi_{i2}}{\xi_{i1}}\biggr\}
+ \frac{\lambda\phi(\lambda\xi_{i2})}{\Phi(\lambda\xi_{i2})}
\biggl\{-\frac{\xi_{i2}}{\alpha} + \frac{c_{\alpha}\xi_{i1}}{2} \biggr\},
\]
\[
c_{i} = -\frac{c_{\alpha}}{2}\biggl\{\xi_{i1}\xi_{i2} - \frac{\xi_{i2}}{\xi_{i1}}\biggr\}
+ \frac{\phi(\lambda\xi_{i2})}{2\Phi(\lambda\xi_{i2})}
(2\xi_{i2} + \lambda c_{\lambda}\xi_{i1}),
\]
for $i=1,\ldots,n$.

\subsection{Response perturbation}\label{resp_pert}

We shall consider here that each $y_i$ is perturbed as $y_{iw} = y_i + \omega_{i}s_{y}$,
where $s_{y}$ is a scale factor that may be estimated by the standard deviation of $\bm{y}$.
In this case, the perturbed log-likelihood function is given by
\[
\ell(\bm{\theta|\bm{\omega}}) = -\frac{n}{2}\log(8\pi) + \sum_{i=1}^{n}\log(\xi_{i1w_{1}})
-\frac{1}{2}\sum_{i=1}^{n}\xi_{i2w_{1}}^{2},
\]
where $\xi_{i1w_{1}}=\xi_{i1w_{1}}(\bm{\theta}) = 2\alpha^{-1}\cosh([y_{iw} -
\bm{x}_{i}^\top\bm{\beta} + c(\alpha,\lambda)]/2)$,
$\xi_{i2w_{1}}=\xi_{i2w_{1}}(\bm{\theta}) = 2\alpha^{-1}\sinh([y_{iw} -
\bm{x}_{i}^\top\bm{\beta} + c(\alpha,\lambda)]/2)$ and
$\bm{\omega}_{0} = (0, \ldots,0)^{\top}$ is the vector of no perturbations.
Here,
\[
\bm{\Delta}_{\bm{\beta}} = s_{y}\bm{X}^{\top}\widehat{\bm{V}},
\qquad
\bm{\Delta}_{\alpha} = s_{y}\widehat{\bm{h}}^\top,
\qquad
\bm{\Delta}_{\lambda} = s_{y}\widehat{\bm{b}}^\top.
\]

\subsection{Explanatory variable perturbation}

Consider now an additive perturbation on a particular continuous explanatory
variable, namely  $\bm{x}_{j}$, by making $x_{ijw} = x_{ij} + \omega_{i}s_{x}$,
where $s_{x}$ is a scale factor that may be estimated by the standard deviation of $\bm{x}_{j}$.
This perturbation scheme leads to the following expression for the log-likelihood function:
\[
\ell(\bm{\theta|\bm{\omega}}) = -\frac{n}{2}\log(8\pi) + \sum_{i=1}^{n}\log(\xi_{i1w_{2}})
-\frac{1}{2}\sum_{i=1}^{n}\xi_{i2w_{2}}^{2},
\]
where $\xi_{i1w_{2}}=\xi_{i1w_{2}}(\bm{\theta}) = 2\alpha^{-1}\cosh([y_{i} -
\bm{x}_{iw}^\top\bm{\beta} + c(\alpha,\lambda)]/2)$,
$\xi_{i2w_{2}}=\xi_{i2w_{2}}(\bm{\theta}) = 2\alpha^{-1}\sinh([y_{i} -
\bm{x}_{iw}^\top\bm{\beta} + c(\alpha,\lambda)]/2)$, with
$\bm{x}_{iw} = (x_{i1},\ldots, x_{ijw},\ldots,x_{ip})^{\top}$.
Here, $\bm{\omega}_{0} = (0, \ldots,0)^{\top}$ is the vector of no perturbations.
Under this perturbation scheme, we have
\[
\bm{\Delta}_{\bm{\beta}} = -s_{x}\widehat{\beta}_{j}\bm{X}^{\top}\widehat{\bm{V}}
+ s_{x}\bm{c}_{j}\widehat{\bm{s}}^\top,
\qquad
\bm{\Delta}_{\alpha} = -s_{x}\widehat{\beta}_{j}\widehat{\bm{h}}^\top,
\qquad
\bm{\Delta}_{\lambda} = -s_{x}\widehat{\beta}_{j}\widehat{\bm{b}}^\top,
\]
where $\bm{c}_{j}$ denotes a $p\times 1$ vector with 1 at the $j$th position and zero elsewhere and
$\widehat{\beta}_{j}$ denotes the $j$th element of $\widehat{\bm{\beta}}$, for $j=1,\ldots,p$.

\section{Generalized leverage}\label{leverage}

In what follows we shall use the generalized leverage proposed by \cite{WeiHuFung1998},
which is defined as
$\bm{GL}(\widetilde{\bm{\theta}}) = \partial\widetilde{\bm{y}}/\partial\bm{y}^{\top}$,
where $\bm{\theta}$ is an $s$-vector such that $\Es(\bm{y}) = \bm{\mu}(\bm{\theta})$ and
$\widetilde{\bm{\theta}}$ is an estimator of $\bm{\theta}$, with
$\widetilde{\bm{y}} = \bm{\mu}(\widetilde{\bm{\theta}})$.
Here, the $(i, l)$ element of $\bm{GL}(\widetilde{\bm{\theta}})$, i.e.~the
generalized leverage of the estimator $\widetilde{\bm{\theta}}$ at
$(i, l)$, is the instantaneous rate of change in $i$th predicted value with respect to the $l$th
response value. As noted by the authors, the generalized leverage is invariant under
reparameterization and observations with large $GL_{ij}$ are leverage points.
\cite{WeiHuFung1998} have shown that the generalized leverage is obtained by
evaluating
\[
\bm{GL}(\bm{\theta}) = \bm{D}_{\bm{\theta}}(-\Ddot{\bm{L}}_{\bm{\theta}\bm{\theta}})^{-1}
\Ddot{\bm{L}}_{\bm{\theta}\bm{y}},
\]
at $\bm{\theta} = \widehat{\bm{\theta}}$, where
$\bm{D}_{\bm{\theta}} = \partial\bm{\mu}/\partial\bm{\theta}^{\top}$
and $\Ddot{\bm{L}}_{\bm{\theta}\bm{y}} =
\partial^2\ell(\bm{\theta})/\partial\bm{\theta}\partial\bm{y}^{\top}$.

After some algebra, we have that
\[
\bm{D}_{\bm{\theta}} =
\begin{bmatrix}
\bm{X} & \bm{0} & \bm{0}
\end{bmatrix}
\qquad{\rm and}\qquad
\Ddot{\bm{L}}_{\bm{\theta}\bm{y}} = -
\begin{bmatrix}
\bm{X}^{\top}\bm{V}\\
\bm{h}^{\top} \\
\bm{b}^\top
\end{bmatrix}.
\]
Thus, from these quantities, we can obtain the generalized leverage.

\section{Application}\label{application}

In this section we shall illustrate the usefulness of the
proposed regression model. The fatigue processes
are by excellence ideally modeled by the Birnbaum--Saunders
distribution due to its genesis. We consider the data set given in \cite{McCool80}
and reported in \cite{Chan-et-al-2008}.
These data consist of times to failure ($T$) in
rolling contact fatigue of ten hardened steel specimens
tested at each of four values of four contact stress ($x$). The data
were obtained using a 4-ball rolling contact test rig at the
Princeton Laboratories of Mobil Research and Development Co.
Similarly to \cite{Chan-et-al-2008}, we consider the following regression model:
\[
y_{i} = \beta_{1} + \beta_{2}\log(x_{i}) + \varepsilon_{i},
\quad i = 1,\ldots,40,
\]
where $y_{i} = \log(T_{i})$ and $\varepsilon_{i}\sim{\rm SSN}(\alpha,-c(\alpha,\lambda),2,\lambda)$,
for $i=1,\ldots,40$. All the computations were
done using the {\tt Ox} matrix programming language \citep{DcK2006}.
{\tt Ox} is freely distributed for academic
purposes and available at {http://www.doornik.com}.

\begin{table}[!htp]
\begin{center}
\caption{Maximum likelihood estimates.}\label{tab1}
\begin{tabular}{l rrcrr}\hline
            & \multicolumn{2}{c}{log-BS}  && \multicolumn{2}{c}{skewed log-BS}\\\cline{2-3}\cline{5-6}
Parameter   &  Estimate &  SE   &&  Estimate &  SE \\\cline{1-3}\cline{5-6}
$\beta_{1}$ &   0.0978  &    0.1707 &&    0.1657  &    0.1759 \\
$\beta_{2}$ & $-14.1164$&    1.5714 &&  $-13.8710$&    1.5887 \\
$\alpha$    &   1.2791  &    0.1438 &&    2.0119  &    0.3487 \\
$\lambda$   &  ---       &  ---       &&    1.6423  &    0.5679 \\ \cline{1-3}\cline{5-6}
Log-likelihood & $-61.62$ &  && $-58.68$  &  \\
AIC            & 129.24 &  && 125.36 &  \\
BIC            & 134.31 &  && 132.12 &  \\
HQIC           & 131.07 &  && 127.80 &  \\\hline
\end{tabular}
\end{center}
\end{table}
Table \ref{tab1} lists the MLEs of the model parameters, asymptotic standard errors (SE),
the values of the log-likelihood functions and the statistics
AIC (Akaike Information Criterion), BIC (Bayesian Information Criterion)
and HQIC (Hannan-Quinn Information Criterion) for
the skewed log-BS and log-BS regression models.
The SE of the estimates for the skewed log-BS model
were obtained using the observed
information matrix given in Section \ref{model}, while the SE of the
estimates for the log-BS model were obtained using the observed information
matrix given, for example, in \cite{Galea-etal-2004}.
The estimatives of $\beta_{1}$ and $\beta_{2}$ differ slightly between the two
models. The skewed log-BS model yields the highest value
of the log-likelihood function and smallest values of the AIC, BIC
and HQIC statistics. From the values of these statistics,
the skewed log-BS model outperforms the BS model and should be prefered.
The likelihood ratio (LR) statistic to the null hypothesis $\lambda=0$
is in accordance with the information criteria (LR = 5.88
and the associated critical level of the $\chi_{1}^{2}$ at 5\% is 3.84).

In what follows, we shall apply the generalized leverage and local influence methods developed
in the previous sections for the purpose of identifying influential observations
in the skewed log-BS regression model fitted to the data set.
\begin{figure}
\centering
\includegraphics[scale=0.4]{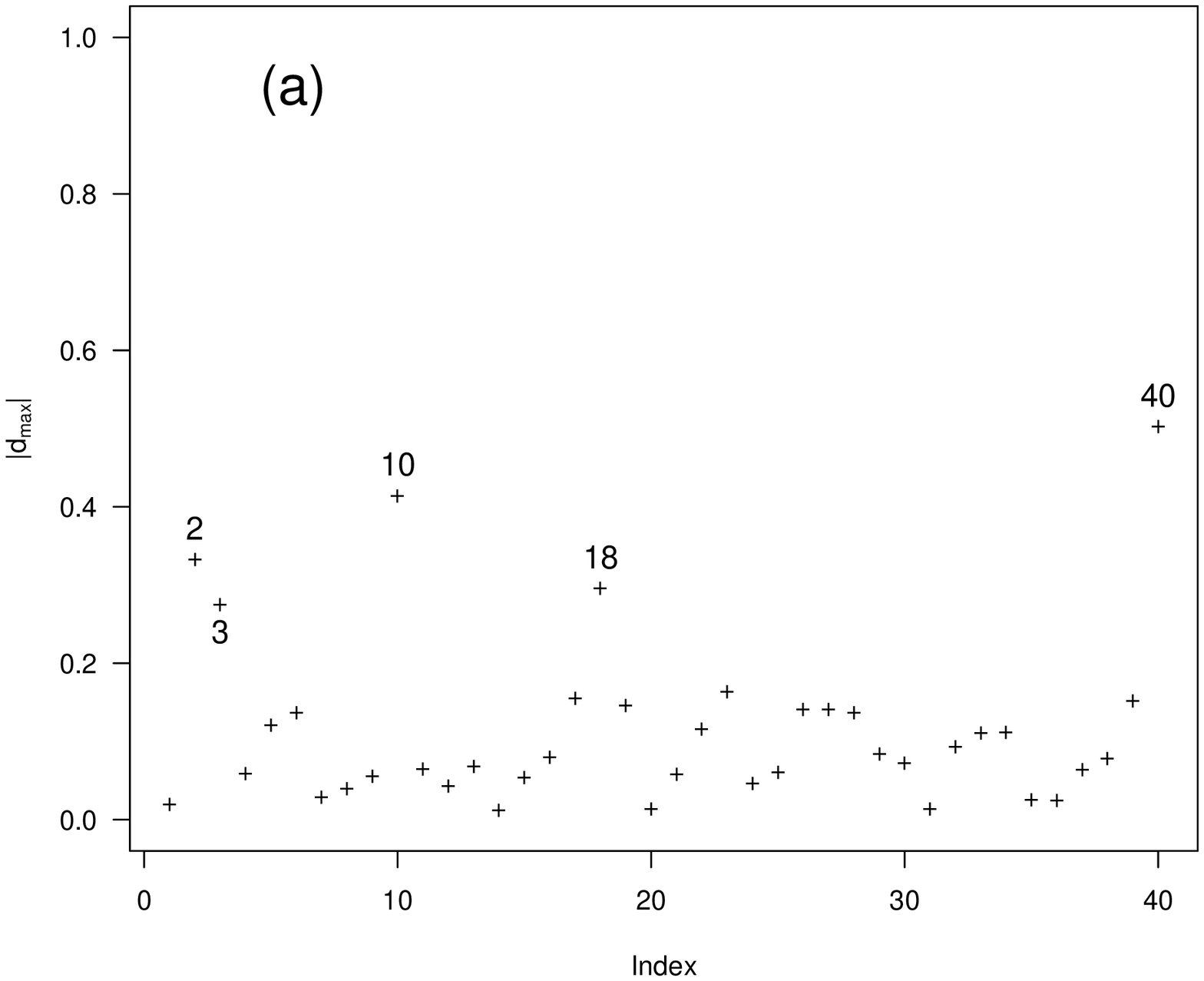}
\includegraphics[scale=0.4]{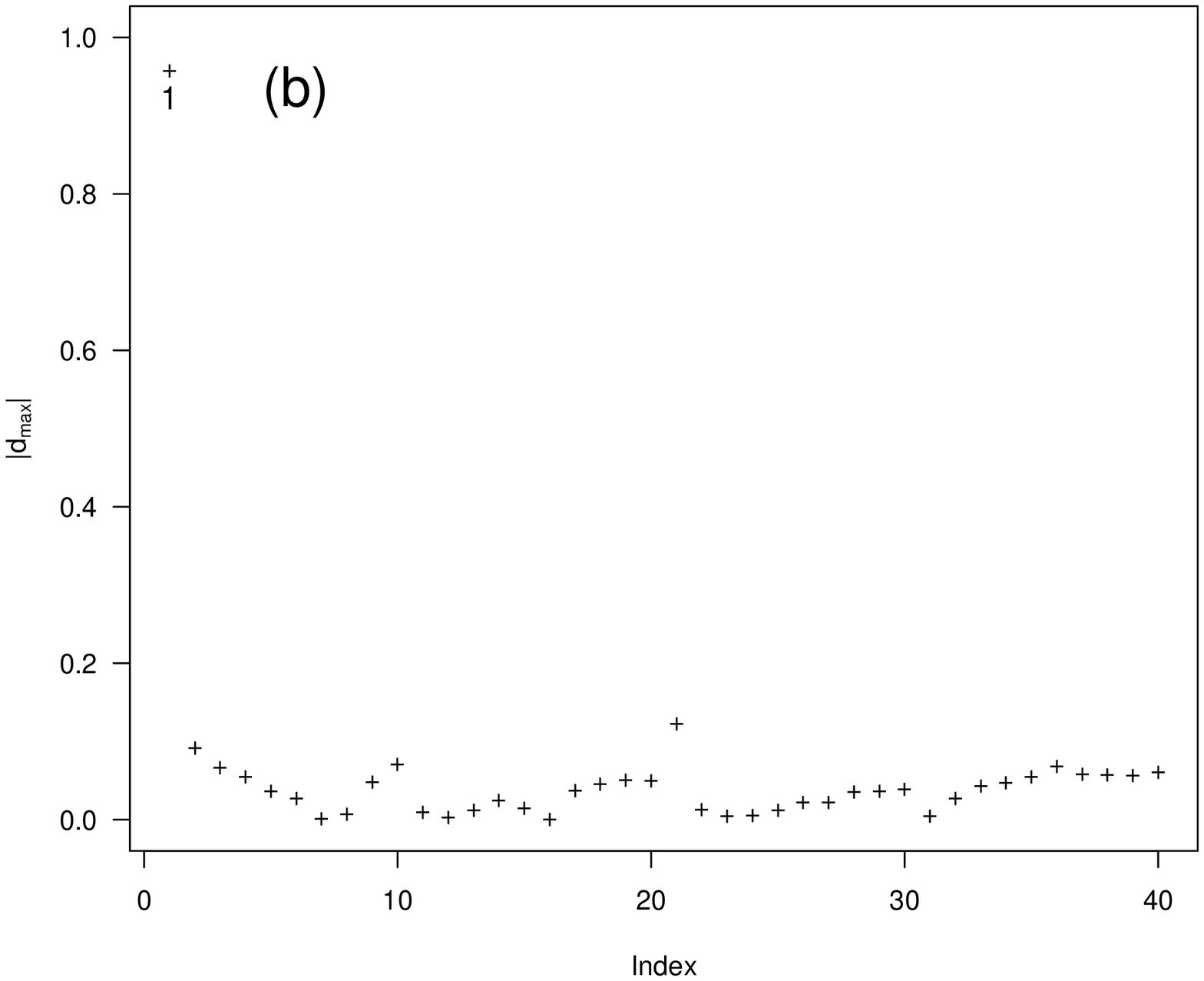}
\includegraphics[scale=0.4]{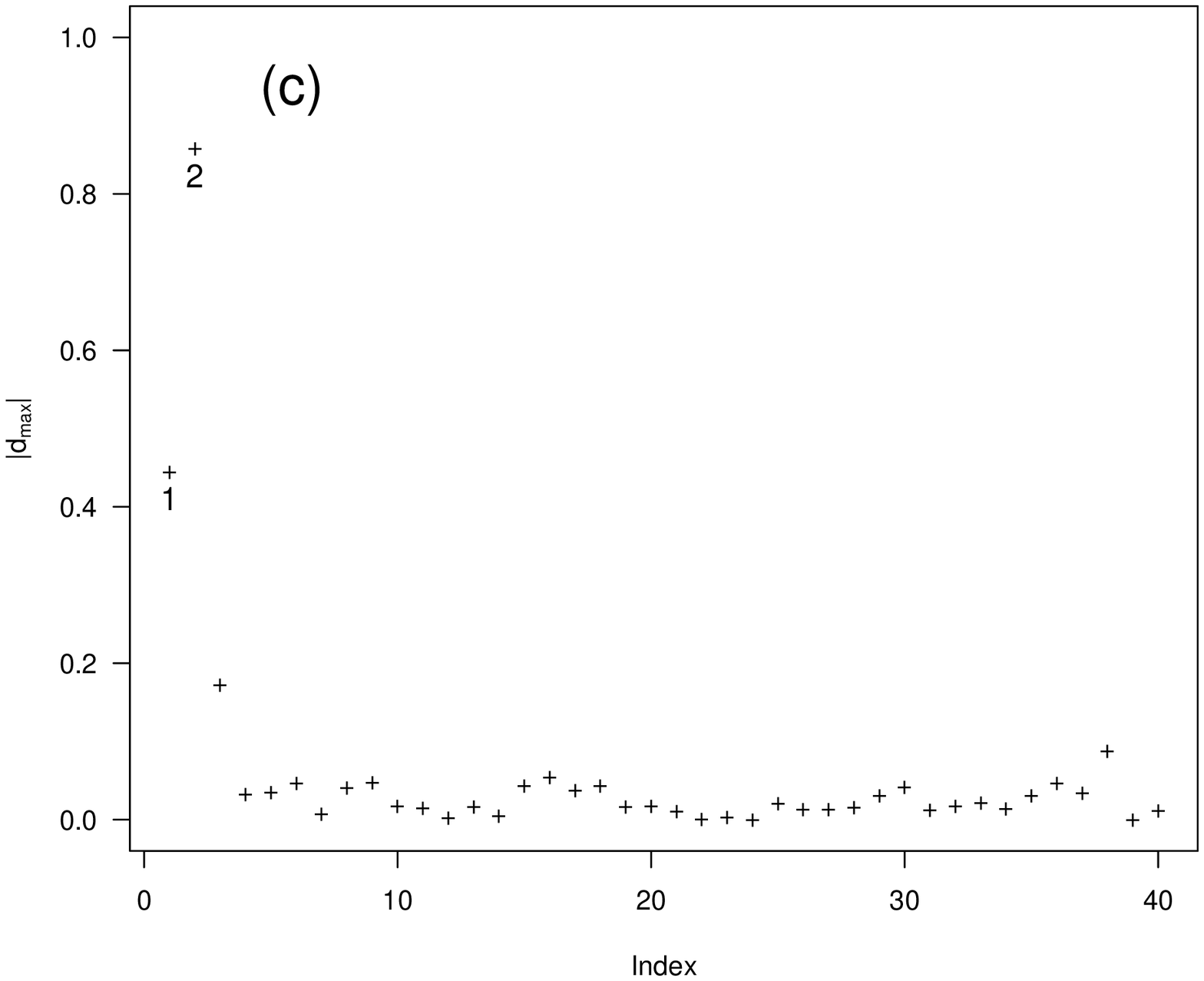}
\includegraphics[scale=0.4]{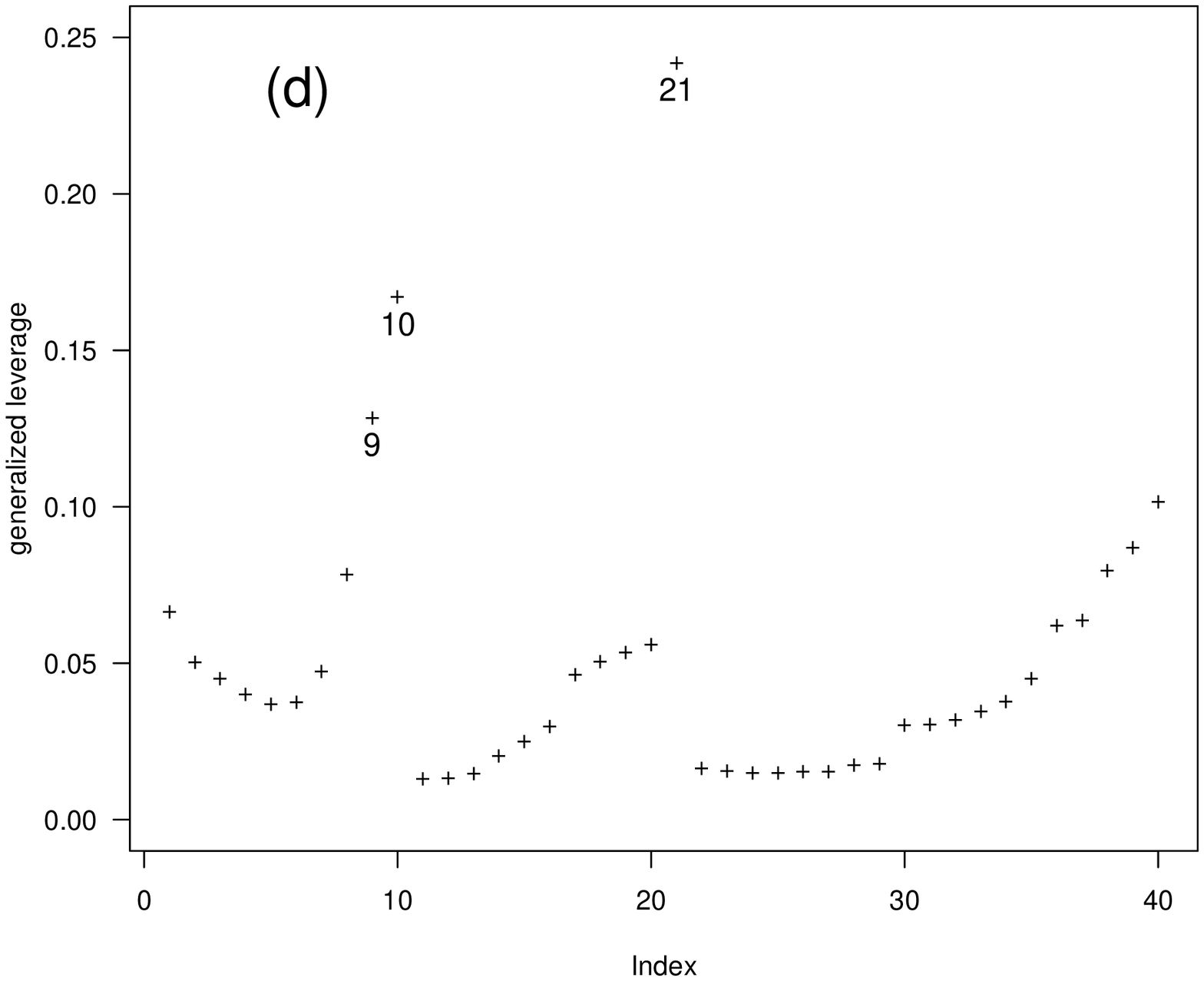}
\caption{Index plots of $|\bm{d}_{\max}|$ for $\widehat{\bm{\theta}}$ under case weighting ({\bf a}),
response ({\bf b}) and covariate ({\bf c})
perturbations, and generalized leverage ({\bf d}).}\label{influence_plot}
\end{figure}
Figure \ref{influence_plot} gives the $|\bm{d}_{\max}|$
corresponding to $\widehat{\bm{\theta}}$ for  different perturbation schemes
and the generalized leverage. An inspection of Figure \ref{influence_plot} reveals that
based on case-weight perturbation (Figure \ref{influence_plot}({\bf a})),
we observed that the cases \#2, \#3, \#10, \#18 and \#40
have more pronounced influence than the other observations.
The case \#1 appears with outstanding influence based on response perturbation
(Figures \ref{influence_plot}({\bf b})).
From the Figure \ref{influence_plot}({\bf c}) (covariate perturbation),
the case \#1 and \#2 have more pronounced influence than the other observations.
Figure \ref{influence_plot}({\bf d})
reveals that the cases \#9, \#10 and \#21 have influence on their own-fitted values.

Based on Figure~\ref{influence_plot}, we eliminated those most influential
observations and refitted the skewed log-BS regression model. In Table \ref{dropping}
we have the relative changes of each parameter estimate,
defined by $\mbox{RC} = |(\widehat{\theta_j} - \widehat{\theta}_{j(i)})/\widehat{\theta}_j|$,
and the corresponding SE,
where $\widehat{\theta}_{j(i)}$ denotes the maximum likelihood
estimate of $\theta_{j}$, after removing the  $i$th observation.
As can be seen, except for the case \#21 corresponding to the parameter $\lambda$,
the relative changes for the maximum likelihood estimates of $\beta_{2}$, $\alpha$ and $\lambda$
are very little pronounced. Also, the significance of these parameters
are not modified in all cases considered. Case \#21
represents the smallest value of the time to failure. Further,
$\beta_{1}$ becomes not significant in all cases considered similar to the
skewed log-BS regression model fitted considering all observations (Table \ref{tab1}).
\begin{table}[!htp]
\begin{center}
\caption{Relative changes dropping the cases indicated, and
the corresponding asymptotic standard errors.}\label{dropping}
\begin{tabular}{l cccccccc}\hline
            & \multicolumn{2}{c}{$\beta_{1}$}  & \multicolumn{2}{c}{$\beta_{2}$}
            & \multicolumn{2}{c}{$\alpha$} & \multicolumn{2}{c}{$\lambda$}\\
Dropping  &  RC & SE   & RC & SE   &  RC & SE & RC & SE \\\hline
\#1           &   0.201  &   0.180  &   0.029  &   1.614  &   0.026  &   0.341  &   0.035  &   0.543  \\
\#2           &   0.161  &   0.180  &   0.024  &   1.618  &   0.013  &   0.345  &   0.030  &   0.544  \\
\#3           &   0.145  &   0.180  &   0.022  &   1.619  &   0.009  &   0.347  &   0.030  &   0.544  \\
\#9           &   0.304  &   0.181  &   0.033  &   1.665  &   0.027  &   0.369  &   0.068  &   0.609  \\
\#10          &   0.543  &   0.180  &   0.051  &   1.675  &   0.005  &   0.358  &   0.044  &   0.588  \\
\#18          &   0.336  &   0.177  &   0.011  &   1.558  &   0.015  &   0.347  &   0.000  &   0.569  \\
\#21          &   0.549  &   0.132  &   0.173  &   1.125  &   0.913  &   0.150  &   3.312  &   0.406  \\
\#40          &   0.121  &   0.176  &   0.027  &   1.626  &   0.014  &   0.346  &   0.019  &   0.565  \\\hline
\end{tabular}
\end{center}
\end{table}

\section{Concluding remarks}\label{conclusion}

In this paper we have introduced a log-Birnbaum--Saunders regression model
with asymmetric errors, extending the usual log-BS regression model.
The random errors of the regression model follow a skewed
sinh-normal distribution, recently derived by \cite{Leiva-et-al-2010-CSTM}.
The estimation of the model parameters is approached by the method
of maximum likelihood and the observed information matrix is derived.
We also consider diagnostic techniques that can be employed to identify influential observations.
Appropriate matrices for assessing local influence on the parameter
estimates under different perturbation schemes are
obtained. The expressions derived are simple, compact
and can be easily implemented into any mathematical or statistical/econometric programming
environment with numerical linear algebra facilities,
such as {\tt R} \citep{R2009} and {\tt Ox} \citep{DcK2006}, among others,
i.e.~our formulas related with this class of regression model are manageable, and with the use of modern
computer resources, may turn into
adequate tools comprising the arsenal of applied statisticians.
Finally, an application to a real data set is presented
to illustrate the usefulness of the proposed model.

As future research, it should be noticed that
some generalizations of the proposed model could be done. For example, a
skewed log-BS regression model that allows us consider censored samples
could be introduced. Following \cite{XiWei07},
one could introduce a skewed log-BS regression model in which the
parameter $\alpha$ is considered different for each observation,
i.e.~to propose an heteroscedastic skewed log-BS regression model.
Also, a skewed log-BS nonlinear regression model could be proposed, and so forth.

\section*{Acknowledgments}

The financial support from FAPESP (Brazil) is gratefully acknowledged.

\appendix
\section*{Appendix}

After extensive algebraic manipulations,
the quantities necessary to obtain the observed information matrix
for the parameter vector $\bm{\theta}=(\bm{\beta}^\top,\alpha,\lambda)^\top$ presented
in the Section \ref{model} are given by
\[
v_{i} = v_{i}(\bm{\theta}) = \frac{1}{4}\biggl\{2\xi_{i2}^{2} + \frac{4}{\alpha^2}
- 1 + \frac{\xi_{i2}^{2}}{\xi_{i1}^{2}} - \frac{\lambda\xi_{i2}\phi(\lambda\xi_{i2})}{\Phi(\lambda\xi_{i2})}
+ \frac{\lambda^3\xi_{i1}^2\xi_{i2}\phi(\lambda\xi_{i2})}{\Phi(\lambda\xi_{i2})}
+ \frac{\lambda^2\xi_{i1}^2\phi(\lambda\xi_{i2})^2}{\Phi(\lambda\xi_{i2})^2}\biggr\},
\]
\begin{align*}
h_{i} = h_{i}(\bm{\theta}) &= \frac{\xi_{i1}\xi_{i2}}{\alpha}
-\frac{c_{\alpha}}{4}\biggl\{2\xi_{i2}^{2} + \frac{4}{\alpha^2}
- 1 + \frac{\xi_{i2}^{2}}{\xi_{i1}^{2}}\biggr\}\\
&+\frac{\lambda c_{\alpha}}{4}\biggl\{\frac{\xi_{i2}\phi(\lambda\xi_{i2})}{\Phi(\lambda\xi_{i2})}
 - \frac{\lambda^2\xi_{i1}^2\xi_{i2}\phi(\lambda\xi_{i2})}{\Phi(\lambda\xi_{i2})}
- \frac{\lambda\xi_{i1}^2\phi(\lambda\xi_{i2})^2}{\Phi(\lambda\xi_{i2})^2}\biggr\}\\
&-\frac{\lambda}{2\alpha}\biggl\{\frac{\xi_{i1}\phi(\lambda\xi_{i2})}{\Phi(\lambda\xi_{i2})}
 - \frac{\lambda^2\xi_{i1}\xi_{i2}^2\phi(\lambda\xi_{i2})}{\Phi(\lambda\xi_{i2})}
- \frac{\lambda\xi_{i1}\xi_{i2}\phi(\lambda\xi_{i2})^2}{\Phi(\lambda\xi_{i2})^2}\biggr\},
\end{align*}
\begin{align*}
b_{i} = b_{i}(\bm{\theta}) &=  -\frac{c_{\lambda}}{4}\biggl\{2\xi_{i2}^{2} + \frac{4}{\alpha^2}
- 1 + \frac{\xi_{i2}^{2}}{\xi_{i1}^{2}}\biggr\}\\
&+\frac{\lambda c_{\lambda}}{4}\biggl\{\frac{\xi_{i2}\phi(\lambda\xi_{i2})}{\Phi(\lambda\xi_{i2})}
 - \frac{\lambda^2\xi_{i1}^2\xi_{i2}\phi(\lambda\xi_{i2})}{\Phi(\lambda\xi_{i2})}
- \frac{\lambda\xi_{i1}^2\phi(\lambda\xi_{i2})^2}{\Phi(\lambda\xi_{i2})^2}\biggr\}\\
&+\frac{1}{2}\biggl\{\frac{\xi_{i1}\phi(\lambda\xi_{i2})}{\Phi(\lambda\xi_{i2})}
 - \frac{\lambda^2\xi_{i1}\xi_{i2}^2\phi(\lambda\xi_{i2})}{\Phi(\lambda\xi_{i2})}
- \frac{\lambda\xi_{i1}\xi_{i2}\phi(\lambda\xi_{i2})^2}{\Phi(\lambda\xi_{i2})^2}\biggr\},
\end{align*}
\begin{align*}
k_{i1} = k_{i1}(\bm{\theta}) &=  \frac{1}{\alpha^2} - \frac{3\xi_{i2}^2}{\alpha^2}
- \frac{c'_{\alpha}}{2}\biggl\{\xi_{i1}\xi_{i2}-\frac{\xi_{i2}}{\xi_{i1}}\biggr\}
+  \frac{\lambda c'_{\alpha}\xi_{i1}\phi(\lambda\xi_{i2})}{2\Phi(\lambda\xi_{i2})}
+\frac{c_{\alpha}\xi_{i1}\xi_{i2}}{\alpha}\\
&- \frac{c'_{\alpha}}{2}\biggl\{-\frac{2\xi_{i1}\xi_{i2}}{\alpha}
+ \frac{c_{\alpha}}{2}(\xi_{i1}^2 + \xi_{i2}^2) - \frac{2c_{\alpha}}{\alpha^2\xi_{i1}^2}\biggr\}
-\frac{\lambda\phi(\lambda\xi_{i2})}{\alpha\Phi(\lambda\xi_{i2})}
\biggl\{-\frac{\xi_{i2}}{\alpha} + \frac{c_{\alpha}\xi_{i1}}{2}\biggr\}\\
&+ \frac{\lambda c_{\alpha}\phi(\lambda\xi_{i2})}{2\Phi(\lambda\xi_{i2})}
\biggl\{-\frac{\xi_{i1}}{\alpha} + \frac{c_{\alpha}\xi_{i2}}{2}\biggr\}
+ \frac{\lambda\xi_{i2}\phi(\lambda\xi_{i2})}{\alpha^2\Phi(\lambda\xi_{i2})}\\
& -\frac{\lambda^2c_{\alpha}\xi_{i1}\phi(\lambda\xi_{i2})}{2\Phi(\lambda\xi_{i2})}
\biggl\{-\frac{\xi_{i2}}{\alpha} + \frac{c_{\alpha}\xi_{i1}}{2} \biggr\}
\biggl\{\lambda\xi_{i2} + \frac{\phi(\lambda\xi_{i2})}{\Phi(\lambda\xi_{i2})} \biggr\}\\
&+\frac{\lambda^2\xi_{i2}\phi(\lambda\xi_{i2})}{\alpha\Phi(\lambda\xi_{i2})}
\biggl\{-\frac{\xi_{i2}}{\alpha} + \frac{c_{\alpha}\xi_{i1}}{2} \biggr\}
\biggl\{\lambda\xi_{i2} + \frac{\phi(\lambda\xi_{i2})}{\Phi(\lambda\xi_{i2})} \biggr\},
\end{align*}
\begin{align*}
k_{i2} = k_{i2}(\bm{\theta}) &=  - \frac{c_{\alpha\lambda}}{2}\biggl\{\xi_{i1}\xi_{i2}-\frac{\xi_{i2}}{\xi_{i1}}\biggr\}
+  \frac{\lambda c_{\alpha\lambda}\xi_{i1}\phi(\lambda\xi_{i2})}{2\Phi(\lambda\xi_{i2})}\\
&- \frac{c_{\lambda}}{2}\biggl\{-\frac{2\xi_{i1}\xi_{i2}}{\alpha}
+ \frac{c_{\alpha}}{2}(\xi_{i1}^2 + \xi_{i2}^2) - \frac{2c_{\alpha}}{\alpha^2\xi_{i1}^2}\biggr\}\\
&+ \frac{\lambda c_{\lambda}\phi(\lambda\xi_{i2})}{2\Phi(\lambda\xi_{i2})}
\biggl\{-\frac{\xi_{i1}}{\alpha} + \frac{c_{\alpha}\xi_{i2}}{2}\biggr\}
+\frac{\phi(\lambda\xi_{i2})}{\Phi(\lambda\xi_{i2})}
\biggl\{-\frac{\xi_{i2}}{\alpha} + \frac{c_{\alpha}\xi_{i1}}{2}\biggr\}\\
& -\frac{\lambda^2c_{\lambda}\xi_{i1}\phi(\lambda\xi_{i2})}{2\Phi(\lambda\xi_{i2})}
\biggl\{-\frac{\xi_{i2}}{\alpha} + \frac{c_{\alpha}\xi_{i1}}{2} \biggr\}
\biggl\{\lambda\xi_{i2} + \frac{\phi(\lambda\xi_{i2})}{\Phi(\lambda\xi_{i2})} \biggr\}\\
&-\frac{\lambda\xi_{i2}\phi(\lambda\xi_{i2})}{\alpha\Phi(\lambda\xi_{i2})}
\biggl\{-\frac{\xi_{i2}}{\alpha} + \frac{c_{\alpha}\xi_{i1}}{2} \biggr\}
\biggl\{\lambda\xi_{i2} + \frac{\phi(\lambda\xi_{i2})}{\Phi(\lambda\xi_{i2})} \biggr\},
\end{align*}
\begin{align*}
k_{i3} = k_{i3}(\bm{\theta}) &=  - \frac{c'_{\lambda}}{2}\biggl\{\xi_{i1}\xi_{i2}-\frac{\xi_{i2}}{\xi_{i1}}\biggr\}
+  \frac{c_{\lambda}\xi_{i1}\phi(\lambda\xi_{i2})}{\Phi(\lambda\xi_{i2})}
+  \frac{\lambda c'_{\lambda}\xi_{i1}\phi(\lambda\xi_{i2})}{2\Phi(\lambda\xi_{i2})}\\
&- \frac{c_{\lambda}^2}{4}\biggl\{2\xi_{i2}^{2} + \frac{4}{\alpha^2}
- 1 + \frac{\xi_{i2}^{2}}{\xi_{i1}^{2}}\biggr\}
+  \frac{\lambda c_{\lambda}^2\xi_{i2}\phi(\lambda\xi_{i2})}{4\Phi(\lambda\xi_{i2})}\\
&-\frac{\lambda^2c_{\lambda}\xi_{i1}\xi_{i2}\phi(\lambda\xi_{i2})}{2\Phi(\lambda\xi_{i2})}
\biggl\{\xi_{i2} + \frac{\lambda c_{\lambda}\xi_{i1}}{2}\biggr\}
-\frac{\lambda c_{\lambda}\xi_{i1}\phi(\lambda\xi_{i2})^2}{2\Phi(\lambda\xi_{i2})^2}
\biggl\{\xi_{i2} + \frac{\lambda c_{\lambda}\xi_{i1}}{2}\biggr\}\\
&-\frac{\lambda\xi_{i2}^2\phi(\lambda\xi_{i2})}{2\Phi(\lambda\xi_{i2})}
\biggl\{\xi_{i2} + \frac{\lambda c_{\lambda}\xi_{i1}}{2}\biggr\}
-\frac{\xi_{i2}\phi(\lambda\xi_{i2})^2}{2\Phi(\lambda\xi_{i2})^2}
\biggl\{\xi_{i2} + \frac{\lambda c_{\lambda}\xi_{i1}}{2}\biggr\},
\end{align*}
for $i=1,\ldots,n$. Also,
\[
c'_{\alpha} = c'_{\alpha}(\alpha,\lambda) =
-4\alpha\int_{-\infty}^{\infty}w^3(4+\alpha^2w^2)^{-3/2}\phi(w)\Phi(\lambda w)dw,
\]
\[
c'_{\lambda} = c'_{\lambda}(\alpha,\lambda) =
-4\lambda\int_{-\infty}^{\infty}w^3\sinh^{-1}(\alpha w/2)\phi(w)\phi(\lambda w)dw,
\]
\[
c_{\alpha\lambda} = c_{\alpha\lambda}(\alpha,\lambda) =
4\int_{-\infty}^{\infty}w^2(4+\alpha^2w^2)^{-1/2}\phi(w)\phi(\lambda w)dw.
\]


{\small
\begin{thebibliography}{99}

\bibitem[Azzaline(1985)]{Azzaline1985}
Azzaline, A. (1985). A class of distributions which includes the normal ones.
{\em Scandinavian Journal of Statistics} {\bf 12}, 171--178.

\bibitem[Balakrishnan et al.(2007)]{Balakrishnan-et-al-2007}
Balakrishnan, N., Leiva, V., L\'opez, J. (2007).
Acceptance sampling plans from truncated life tests from generalized Birnbaum--Saunders
distribution. {\em Communications in Statistics -- Simulation and Computation} {\bf 36}, 643--656.

\bibitem[Bhatti(2010)]{Bhatti2010}
Bhatti, C.R. (2010). The Birnbaum--Saunders autoregressive conditional duration model.
{\em Mathematics and Computers in Simulation} {\bf 80}, 2062--2078.

\bibitem[Birnbaum and Saunders(1969a)]{BSa1969a}
Birnbaum, Z.W., Saunders, S.C. (1969a).
\newblock A new family of life distributions.
\newblock {\em Journal of Applied Probability\/} {\bf 6}, 319--327.

\bibitem[Birnbaum and Saunders(1969b)]{BSa1969b}
Birnbaum, Z.W., Saunders, S.C. (1969b).
\newblock Estimation for a family of life distributions with applications to fatigue.
\newblock {\em Journal of Applied Probability\/} {\bf 6}, 328--377.

\bibitem[Cancho et al.(2010)]{Cancho-et-al-2010}
Cancho, V.G., Ortega, E.E.M., Paula, G.A. (2010).
On estimation and influence diagnostics for log-Birnbaum--Saunders
Student-$t$ regression models: Full Bayesian analysis.
{\em Journal of Statistical Planning and Inference\/} {\bf 140}, 2486--2496.

\bibitem[Chan et al.(2008)]{Chan-et-al-2008}
Chan, P.S., Ng, H.K.T., Balakrishnan, N., Zhou, Q. (2008).
Point and interval estimation for extreme-value regression model
under Type-II censoring. {\em Computational Statistics ans Data Analysis}
{\bf 52}, 4040--4058.

\bibitem[Castillo et al.(2009)]{Castillo-et-al-2009}
Castillo, N.O., G\'omez, H.W., Bolfarine, H. (2009).
Epsilon Birnbaum--Saunders distribution family: properties and inference.
{\em Statistical Papers}. {\tt DOI:10.1007/s00362-009-0293-x}.

\bibitem[Cook(1986)]{Cook1986}
Cook, R.D. (1986). Assessment of local influence (with discussion).
\emph{Journal of the Royal Statistical Society B} \textbf{48}, 133--169.

\bibitem[Cox and Hinkley(1974)]{CoxHinkley1974}
Cox, D.R., Hinkley, D.V. (1974). \emph{Theoretical Statistics}.
London: Chapman and Hall.

\bibitem[Desmond(1985)]{Desmond1985}
Desmond, A.F. (1985). Stochastic models of failure in random environments.
{\em Canadian Journal of Statistics\/} {\bf 13}, 171--183.

\bibitem[Desmond(1986)]{Desmond1986}
Desmond, A.F. (1986).
On the relationship between two fatigue-life models.
{\em IEEE Transactions on Reliability\/} {\bf 35}, 167--169.

\bibitem[Desmond et al.(2008)]{Desmond-et-al-2008}
Desmond, A.F., Rodr\'iguez--Yam, G.A., Lu, X. (2008).
Estimation of parameters for a Birnbaum--Saunders regression model with censored data.
{\em Journal of Statistical Computation and Simulation} {\bf 78}, 983--997.

\bibitem[D\'iaz--Garc\'ia and Leiva(2005)]{Diaz-Leiva05}
D\'iaz--Garc\'ia, J.A., Leiva, V. (2005).
\newblock A new family of life distributions based on the elliptically contoured distributions.
\newblock {\em Journal of Statistical Planning and Inference\/} {\bf 128}, 445--457.

\bibitem[Doornik(2006)]{DcK2006}
Doornik, J.A. (2006). {\em An Object-Oriented Matrix Language -- Ox 4\/}, 5th ed.
Timberlake Consultants Press, London.

\bibitem[Galea et al.(2004)]{Galea-etal-2004}
Galea, M., Leiva, V., Paula, G.A.~(2004).
Influence diagnostics in log-Birnbaum--Saunders regression models.
\emph{Journal of Applied Statistics} \textbf{31}, 1049--1064.

\bibitem[G\'omes et al.(2009)]{GPB09}
G\'omes, H.W., Olivares--Pacheco, J.F., Bolfarine, H. (2009).
\newblock An extension of the generalized Birnbaum--Saunders distribution.
\newblock {\em Statistics and Probability Letters\/} {\bf 79}, 331--338.

\bibitem[Guiraud et al.(2009)]{guiraud-et-al-2009}
Guiraud, P., Leiva, V., Fierro, R. (2009).
\newblock A non-central version of the Birnbaum--Saunders distribution for reliability analysis.
\newblock {\em IEEE Transactions on Reliability\/} {\bf 58}, 152--160.

\bibitem[Kundu et al.(2008)]{kundu-et-al-2008}
Kundu, D., Kannan, N., Balakrishnan, N. (2008).
\newblock On the function of Birnbaum--Saunders distribution and associated inference.
\newblock {\em Computational Statistics and Data Analysis\/} {\bf 52}, 2692--2702.

\bibitem[Leiva et al.(2007)]{LBPG2007}
Leiva, V., Barros, M.K., Paula, G.A., Galea, M. (2007).
Influence diagnostics in log-Birnbaum--Saunders regression models
with censored data. {\em Computational Statistics and Data Analysis\/}, {\bf 51},
  5694--5707.

\bibitem[Leiva et al.(2008)]{Leiva-et-al-2008}
Leiva, V., Barros, M., Paula, G.A., Sanhueza, A. (2008).
Generalized Birnbaum--Saunders distributions applied to air
pollutant concentration. {\em Environmetrics}  {\bf 19}, 235--249 .

\bibitem[Leiva et al.(2009)]{Leiva-et-al-2009}
Leiva, V., Sanhueza, A., Angulo, J.M. (2009).
A length-biased version of the Birnbaum--Saunders distribution with application in water quality.
{\em Stochastic Environmental Research and Risk Assessment}  {\bf 23}, 299--307.

\bibitem[Leiva et al.(2010)]{Leiva-et-al-2010-CSTM}
Leiva, V., Vilca, F., Balakrishnan, N., Sanhueza, A. (2010).
A skewed sinh-normal distribution and its properties and application to
air pollution. {\em Communications in Statistics -- Theory and Methods}  {\bf 39}, 426--443.

\bibitem[Lemonte and Cordeiro(2009)]{LemCord09}
Lemonte, A.J., Cordeiro, G.M. (2009).
\newblock Birnbaum--Saunders nonlinear regression models.
\newblock {\em Computational Statistics and Data Analysis\/}
{\bf 53}, 4441--4452.

\bibitem[Lemonte et al.(2007)]{LCNV07}
Lemonte, A.J., Cribari--Neto, F., Vasconcellos, K.L.P. (2007).
\newblock Improved statistical inference for the two-parameter
  Birnbaum--Saunders distribution.
\newblock {\em Computational Statistics and Data Analysis\/} {\bf 51}, 4656--4681.

\bibitem[Lemonte et al.(2010)]{Lemonte-et-al-2010}
Lemonte, A.J., Ferrari, S.L.P., Cribari--Neto, F. (2010).
Improved likelihood inference in Birnbaum--Saunders regressions.
{\em Computational Statistics and Data Analysis\/} {\bf 54}, 1307--1316.

\bibitem[Lemonte et al.(2008)]{LSCN08}
Lemonte, A.J., Simas, A.B., Cribari--Neto, F. (2008).
\newblock Bootstrap-based improved estimators for the two-parameter
  Birnbaum--Saunders distribution.
\newblock {\em Journal of Statistical Computation and Simulation\/} {\bf 78}, 37--49.

\bibitem[Lepadatu et al.(2005)]{Lepadatu-et-al-2005}
Lepadatu, D., Kobi, A., Hambli, R. and Barreau, A. (2005).
\newblock Lifetime multiple response optimization of metal extrusion die.
\newblock {\em Proceedings of the Annual Reliability and Maintainability Symposium\/},
37--42.

\bibitem[Marshall and Olkin(2007)]{MO2007}
Marshall, A.W., Olkin, I. (2007). {\em Life Distributions}. Springer: New York.

\bibitem[McCool(1980)]{McCool80}
McCool, J.I. (1980). Confidence limits for Weibull regression with censored data.
{\em IEEE Transactions on Reliability} {\bf 29}, 145--150.

\bibitem[Nocedal and Wright(1999)]{NocedalWright1999}
Nocedal, J., Wright, S.J. (1999). {\em Numerical Optimization}. Springer: New York.

\bibitem[Press et al.(2007)]{Press-et-al-2007}
Press, W.H., Teulosky, S.A., Vetterling, W.T., Flannery, B.P. (2007).
{\em Numerical Recipes in C: The Art of Scientific Computing\/}, 3rd ed.
Cambridge University Press.

\bibitem[R Development Core Team(2009)]{R2009}
R Development Core Team (2009). {\em R: A Language and Environment for Statistical Computing}.
Vienna, Austria.

\bibitem[Rieck(1989)]{Rieck89}
Rieck, J.R. (1989). {\em Statistical Analysis for the Birnbaum--Saunders Fatigue Life
  Distribution\/}. Ph.D.~dissertation, Clemson University.

\bibitem[Rieck and Nedelman(1991)]{RiekNedelman91}
Rieck, J.R., Nedelman, J.R. (1991).
\newblock A log-linear model for the Birnbaum--Saunders distribution.
\newblock {\em Technometrics\/} {\bf 33}, 51--60.

\bibitem[Saunders(1974)]{Saunders1974}
Saunders, S.C. (1974). A family of random variables closed under reciprocation.
\newblock {\em Journal of the American Statistical Association\/} {\bf 69}, 533--539.

\bibitem[Tisionas(2001)]{Tisionas01}
Tisionas, E.G. (2001).
Bayesian inference in Birnbaum--Saunders regression.
{\em Communications in Statistics -- Theory and Methods\/} {\bf 30},
  179--193.

\bibitem[Xi and Wei(2007)]{XiWei07}
Xi, F.C., Wei, B.C. (2007).
Diagnostics analysis for log-Birnbaum--Saunders regression models.
{\em Computational Statistics and Data Analysis\/} {\bf 51}, 4692--4706.

\bibitem[Xiao et al.(2010)]{Xiao-et-al-2010}
Xiao, Q., Liu, Z., Balakrishnan, N., Lu, X. (2010).
Estimation of the Birnbaum--Saunders regression model with current status data.
{\em Computational Statistics and Data Analysis} {\bf 54}, 326--332.

\bibitem[Xu and Tang(2010)]{XiTang10}
Xu, A., Tang, Y. (2010).
\newblock Reference analysis for Birnbaum--Saunders distribution.
\newblock {\em Computational Statistics and Data Analysis\/} {\bf 54}, 185--192.

\bibitem[Wei et al.(1998)]{WeiHuFung1998}
Wei, B.C., Hu, Y.Q., Fung, W.K.~(1998).
Generalized leverage and its applications. \emph{Scandinavian
Journal of Statistics} \textbf{25}, 25--37.

\bibitem[Wu and Wong(2004)]{WuWong2004}
Wu, J., Wong, A.C.M. (2004). Improved interval estimation for the
two-parameter Birnbaum--Saunders distribution.
{\em Computational Statistics and Data Analysis\/} {\bf 47}, 809--821.
\end{thebibliography}
}
\end{document}